\documentclass[12pt]{iopart}

\usepackage{graphicx}

\usepackage[]{cite}

\expandafter\let\csname equation*\endcsname\relax
\expandafter\let\csname endequation*\endcsname\relax

\usepackage{amsmath}
\usepackage[]{amssymb}
\usepackage[]{mathrsfs}
\usepackage[]{eucal}
\usepackage[]{tensor}
\usepackage[]{amsthm}
\usepackage[]{bm}

\newcommand{\pf}[1]{\mathbf{#1}}
\newcommand{\dd}{\partial}
\newcommand{\hdg}{\star} 
\newcommand{\df}{\mathrm{d}}
\newcommand{\w}{\wedge}

\newcommand{\Lie}{\pounds}
\newcommand{\nab}[1]{\nabla_{\!#1}}

\newcommand{\qqd}{\ , \quad}
\newcommand{\bc}{\begin{center}}
\newcommand{\ec}{\end{center}}
\newcommand{\be}{\begin{equation}}
\newcommand{\ee}{\end{equation}}

\newcommand{\F}{\pf{F}}
\newcommand{\Z}{\pf{Z}}

\newcommand{\FF}{\mathcal{F}}
\newcommand{\GG}{\mathcal{G}}
\newcommand{\HH}{\mathcal{H}}
\newcommand{\LL}{\mathscr{L}}

\newcommand{\norm}[1]{\left\lVert #1 \right\rVert}

\newcommand{\defeq}{\mathrel{\mathop:}=}

\usepackage{dsfont}

\newcommand{\rr}{\mathds{R}}

\usepackage[unicode]{hyperref}
\usepackage{xcolor}
\definecolor{pastgreen}{HTML}{669900}
\definecolor{pastblue}{HTML}{336699}
\definecolor{pastred}{HTML}{990000}
\definecolor{linkcol}{HTML}{663333}
\hypersetup{colorlinks,linkcolor={pastblue},citecolor={pastblue},urlcolor={pastblue}}

\theoremstyle{plain} \newtheorem{tm}{Theorem}[section]
\theoremstyle{plain} \newtheorem{lm}[tm]{Lemma}
\theoremstyle{plain} \newtheorem{defn}[tm]{Definition}
\newcommand{\btm}{\begin{tm}}
\newcommand{\etm}{\end{tm}}
\newcommand{\blm}{\begin{lm}}
\newcommand{\elm}{\end{lm}}
\newcommand{\bdefn}{\begin{defn}}
\newcommand{\edefn}{\end{defn}}

\begin{document}

\begin{flushright}
\texttt{ZTF-EP-24-03}

\texttt{RBI-ThPhys-2024-04}
\end{flushright}

\title[Hexadecapole at the heart of nonlinear electromagnetic fields]{Hexadecapole at the heart of nonlinear electromagnetic fields}

\author{Ana Bokuli\'c$^a$, Tajron Juri\'c$^b$ and Ivica Smoli\'c$^a$}
\address{$^a$ Department of Physics, Faculty of Science, University of Zagreb, Bijeni\v cka cesta 32, 10000 Zagreb, Croatia}
\address{$^b$ Rudjer Bo\v skovi\'c Institute, Bijeni\v cka cesta 54, HR-10002 Zagreb, Croatia}
\eads{\mailto{abokulic@phy.hr}, \mailto{tjuric@irb.hr}, \mailto{ismolic@phy.hr}}

\date{\today}

\begin{abstract}
In classical Maxwell's electromagnetism, monopole term of the electric field is proportional to $r^{-2}$, while higher order multipole terms, sourced by anisotropic sources, fall-off faster. However, in nonlinear electromagnetism even a spherically symmetric field has multipole-like contributions. We prove that the leading sub\-dominant term of the electric field, defined by nonlinear electromagnetic Lagrangian obeying Maxwellian weak field limit, in a static, spherically symmetric, asymptotically flat spacetime, is of the order $O(r^{-6})$ as $r \to \infty$. Moreover, using Lagrange inversion theorem and Fa\`a di Bruno's formula, we derive the series expansion of the electric field from the Taylor series of an analytic electromagnetic Lagrangian.
\end{abstract}

\vspace{2pc}


\section{Introduction} 

Harmonic functions, solutions of the Laplace equation $\Delta u = 0$ on some open set $\Omega \subseteq \rr^m$, are ubiquitous in physics. The most familiar examples are electric scalar potential and stationary distribution of the temperature in the source-free region. Even from a mathematical perspective, harmonic functions stand out with numerous exceptional properties \cite{ABR}: they attain extrema only on the boundary of the domain, their value at each point $s \in \Omega$ is equal to the average value over any ball $B(s,r) \Subset \Omega$, and even if we assume that a harmonic function belongs ``just'' to class $C^2$ of differentiability, it turns out that it is necessarily analytic in the interior of its domain. In case when $m \ge 3$, asymptotic behaviour of harmonic functions becomes particularly restricted: the mere assumption that $u(x) \to 0$ as $\norm{x}\to\infty$ in the complement of a compact set, implies that $u$ is in fact of order $O(\norm{x}^{2-m})$ as $\norm{x}\to\infty$. Physical intuition about the latter result comes from the multipole expansion, in which the monopole term has the slowest fall-off rate at infinity. Moreover, in the case of spherically symmetric solution, the only harmonic function vanishing at infinity is proportional to $\norm{x}^{2-m}$.

\smallskip

All this becomes far less trivial once we move from the linear Laplace equation to its nonlinear generalizations, most notably in the context of electromagnetism. Namely, we already know, both from high-energy experiments and quantum field theoretic predictions, that Maxwell's electromagnetism is not the complete description of the electromagnetic fields in the nature and one needs to look at its extensions, so-called nonlinear electromagnetic (NLE) theories. A prominent family of NLE theories, defined by a Lagrangian density which is a function of two electromagnetic invariants\footnote{The Hodge dual ${\hdg\bm{\omega}}$ of a $p$-form $\bm{\omega}$ is defined as $(4-p)$-form ${\hdg\omega}_{a_{p+1} \dots a_4} = \frac{1}{p!} \, \omega_{b_1 \dots b_p} \tensor{\epsilon}{^{b_1}^{\dots}^{b_p}_{a_{p+1}}_{\dots}_{a_4}}$.}, $\FF \defeq F_{ab} F^{ab}$ and $\GG \defeq F_{ab} \,{\hdg F}^{ab}$, has its roots in the early days of quantum field theory, with Born--Infeld \cite{Born34,BI34} and Euler--Heisenberg Lagrangians \cite{HE36,Dunne04}. Over the past several decades investigation of gravitating NLE fields has intensified \cite{Sorokin21,Bronnikov22rev}, motivated to some extent by a prospect of regularization of spacetime singularities \cite{CCZ18,SZ22,Maeda22,LYGM23}.

\smallskip

In this paper we shall investigate properties of the electric field and the corresponding electric scalar potential at spatial infinity of asymptotically flat, spherically symmetric spacetime, assuming that NLE theory reduces to Maxwell's electromagnetism for weak fields. Here it is convenient to introduce abbreviations for partial derivatives, e.g.~$\LL_\FF \defeq \dd_\FF \LL$ and $\LL_\GG \defeq \dd_\GG \LL$, and $\HH \defeq \sqrt{\FF^2 + \GG^2}$ for the ``radial coordinate'' in the $\FF$-$\GG$ plane. Generalized source-free Maxwell's equations, corresponding to the NLE Lagrangian $\LL(\FF,\GG)$, may be written as
\be\label{eq:gMax}
\df\F = 0 \qqd \df{\hdg\Z} = 0 \, ,
\ee
where $\Z \defeq -4(\LL_\FF \F + \LL_\GG {\hdg\F})$ is an auxiliary 2-form. Finally, in order to make the notion of weak field limit more precise, we rely on the following definition.

\smallskip

\bdefn
We say that an NLE Lagrangian density $\LL(\FF,\GG)$ satisfies the Max\-well\-ian weak field (MWF) limit if it is of $C^1$ class on some neighbourhood of the origin $(0,0)$ and
\be
\LL_\FF(\FF,\GG) = -\frac{1}{4} + O(\HH) \quad \textrm{and} \quad \LL_\GG(\FF,\GG) = O(\HH)
\ee
as $\HH \to 0$.
\edefn

\smallskip

Essential parts of our analysis do not depend on gravitational field equations, so the conclusions may be equally applied to gravitating electromagnetic fields, as well as test fields on a fixed background.

\section{First order correction to Maxwell} 

Let $(M,g_{ab})$ be a smooth 4-dimensional static, spherically symmetric Lorentzian manifold, with the corresponding Killing vector fields, $k^a$ (time translation) and $\{X_1^a,X_2^a,X_3^a\}$ (spherical symmetry), such that $\Lie_k X_i^a = 0$ for all $i$. In canonical choice of coordinates $(t,r,\theta,\varphi)$, under the assumption that $\nab{a} r \ne 0$, the metric may be written in a form 
\be
\df s^2 = -\alpha(r) \, \df t^2 + \beta(r) \, \df r^2 + r^2 (\df\theta^2 + \sin^2\theta \, \df\varphi^2) \, .
\ee
Furthermore, let $\F$ be the electromagnetic field, a solution of (\ref{eq:gMax}), inheriting all the spacetime symmetries, $\Lie_k \F = \Lie_{X_i} \F = 0$ for all $i$. If we introduce the electric 1-form $\pf{E} = -i_k \F$ and the magnetic 1-form $\pf{H} = i_k {\hdg\Z}$, generalized Maxwell's equations (\ref{eq:gMax}) imply that they are closed forms,
\begin{align}
\df\pf{E} = -\df i_k\F = (-\Lie_k + i_k \df) \F & = 0 \, , \\
\df\pf{H} = \df i_k {\hdg\Z} = (\Lie_k - i_k \df) {\hdg\Z} & = 0 \, .
\end{align}
Thus, by Poincar\'e lemma, we know that at least locally we may introduce the electric scalar potential $\Phi$ and the magnetic scalar potential $\Psi$ via
\be
\pf{E} = -\df\Phi \qqd \pf{H} = -\df\Psi \, .
\ee
With another magnetic 1-form $\pf{B} \defeq i_k {\hdg\F}$, which is not necessarily closed, we have a decomposition
\be\label{eq:Fdecomp1}
\alpha(r) \, \F = \pf{k} \w \pf{E} + {\hdg(\pf{k} \w \pf{B})}
\ee
at our disposal. In fact, we may infer even more about the electromagnetic 2-form $\F$, by adapting well-known tools \cite{Heusler} to NLE fields (cf.~remarks in \cite{BS23}).

\blm
Under the assumptions given above, $\pf{E} = E_r \, \df r$ and, at each point where $\LL_\FF \ne 0$, $\pf{B} = B_r \, \df r$. Furthermore, at each point where $\alpha\beta \ne 0$ we have
\be\label{eq:Fdecomp2}
\F = -E_r \, \df t \w \df r + \frac{B_r \, r^2 \sin\theta}{\sqrt{\alpha\beta}} \, \df\theta \w \df\varphi \, .
\ee
\elm

\noindent
\emph{Proof}. Let $K^a$ and $L^a$ be two commuting Killing vector fields and $\bm{\omega}$ a closed 2-form, such that $\Lie_K \bm{\omega} = 0 = \Lie_L \bm{\omega}$. Then, identity $\df i_X i_Y = i_X i_Y \df + i_{[X,Y]} - i_X \Lie_Y + i_Y \Lie_X$ applied with $X^a = K^a$ and $Y^a = L^a$ to $\bm{\omega}$, immediately implies that $\df (i_K i_L \bm{\omega}) = 0$, that is $i_K i_L \bm{\omega}$ is a constant. More concretely, using $K^a = k^a$, $L^a = X_i^a$ and $\bm{\omega} = \F$, we know that $i_k i_{X_i} \F$ is constant for all $i$ and zero on each connected component of spacetime $M$ whose boundary intersects axis on which the corresponding Killing vector field $X_i^a$ vanishes. As $\Lie_Y \Phi = i_Y \df\Phi = -i_Y \pf{E} = i_Y i_k \F = 0$ for any $Y^a \in \{k^a,X_1^a, X_2^a, X_3^a\}$, it follows that $\Phi = \Phi(r)$ and $\pf{E} = -\Phi'(r) \, \df r$. By completely analogous procedure, applied to $\bm{\omega} = {\hdg\Z}$, we deduce that $\Psi = \Psi(r)$ and $\pf{H} = -\Psi'(r) \, \df r$. On the other hand, $\pf{H} = -4(\LL_\FF \pf{B} + \LL_\GG \pf{E})$. Thus, on each point where $\LL_\FF \ne 0$ it follows that $\pf{B} = B_r \, \df r$. Finally, taking into account $\pf{k} = -\alpha(r) \, \df t$ and the Hodge dual
\be
{\hdg(\df t \w \df r)} = -\frac{r^2 \sin\theta}{\sqrt{\alpha\beta}} \, \df\theta \w \df\varphi \, ,
\ee
the decomposition (\ref{eq:Fdecomp2}) immediately follows from (\ref{eq:Fdecomp1}).
\qed

\smallskip

As was already noted by Jacobson \cite{Jacobson07}, product $\alpha\beta$ may be sometimes additionally simplified. For example, NLE energy-momentum tensor \cite{BJS21},
\be
T_{ab} = -4\LL_\FF \widetilde{T}_{ab} + \frac{1}{4} \, \tensor{T}{^c_c} \, g_{ab} \, , \ \textrm{with} \quad \widetilde{T}_{ab} = \frac{1}{4\pi} \left( F_{ac} \tensor{F}{_b^c} - \frac{1}{4} \, \FF g_{ab}\right) ,
\ee
in the spacetime defined above satisfies $\beta T_{tt} + \alpha T_{rr} = 0$ and, as the corresponding combination of Ricci tensor components is reduced to $\beta R_{tt} + \alpha R_{rr} = (\alpha\beta)'/(r\beta)$, in the special case of Einstein's gravitational field equations it follows that the product $\alpha\beta$ is constant (moreover, by rescaling of the time coordinate we may choose $\alpha\beta = 1$). However, as this simplification is not necessary for our analysis, we shall keep the product $\alpha\beta$ unconstrained (up to the assumptions about asymptotic flatness).

\smallskip

Following the idea from \cite{BJS22b}, we introduce auxiliary, ``normalized'' 1-forms
\be
\widehat{\pf{E}} \defeq \frac{1}{\sqrt{\alpha\beta}} \, \pf{E} \qqd \widehat{\pf{B}} \defeq \frac{1}{\sqrt{\alpha\beta}} \, \pf{B} \, ,
\ee
at each point where $\alpha\beta \ne 0$. With this notation electromagnetic invariants may be written as
\be
\FF = 2(\widehat{B}_r^2 - \widehat{E}_r^2) \qqd \GG = 4\widehat{E}_r \widehat{B}_r \, .
\ee
A crucial detail here is that, by an elementary inequality,
\be
\HH = 2(\widehat{E}_r^2 + \widehat{B}_r^2) \le 2(|\widehat{E}_r| + |\widehat{B}_r|)^2 \, .
\ee
Finally, generalized Maxwell's equations are reduced to
\be
\widehat{B}_r = \frac{P}{r^2} \qqd \LL_\FF \widehat{E}_r - \LL_\GG \widehat{B}_r = -\frac{Q}{4r^2} \, ,
\ee
where $Q$ and $P$ are, respectively, electric and magnetic charges, defined via Komar integrals \cite{BJS21},
\be
Q \defeq \oint_\infty {\hdg\Z} \qqd P \defeq \oint_\infty \F \, .
\ee
Now we may state and prove our first result.

\smallskip

\btm
Let $(M,g_{ab})$ be a static, spherically symmetric, asymptotically flat spacetime with electromagnetic field $\F$ which decreases in a sense that $\lim_{r\to\infty} \FF = 0$ and $\lim_{r\to\infty} \GG = 0$. Then, given that there is a radius $r_e > 0$ such that at least in the region $r > r_e$ the electromagnetic field (a) inherits all symmetries and (b) satisfies source-free generalized Maxwell field equations (\ref{eq:gMax}) with NLE Lagrangian density $\LL(\FF,\GG)$ satisfying MWF limit, the electric field has a form $\widehat{E}_r = Qr^{-2} + O(r^{-6})$ as $r \to \infty$.
\etm

\smallskip

\noindent
\emph{Proof}. Asymptotic flatness imposes $\lim_{r\to\infty} \alpha(r) = 1$ and $\lim_{r\to\infty} \beta(r) = 1$. In other words, for any $0 < \epsilon < 1$ there is a radius $r_0 \ge r_e$, such that $|\alpha(r) - 1| \le \epsilon$ and $|\beta(r) - 1| \le \epsilon$ for all $r \ge r_0$ and here we have well-defined 1-forms $\widehat{\pf{E}}$ and $\widehat{\pf{B}}$. Furthermore, MWF limit implies that there are positive constants $\delta,C_1,C_2 > 0$ such that $|\HH| \le \delta$ implies
\be\label{eq:MWF}
\big| \LL_\FF + \frac{1}{4} \big| \le C_1 \HH \quad \textrm{and} \quad \big| \LL_\GG \big| \le C_2 \HH \, .
\ee
On the other hand, asymptotic decrease of electromagnetic invariants implies that for any $\delta > 0$ there is radius $r_1 \ge r_0$ such that $|\HH| \le \delta$ for all $r \ge r_1$. In particular, conditions (\ref{eq:MWF}) and consequently $\LL_\FF \ne 0$ hold for $r \ge r_1$ (assuring, among other things, that conditions of lemma 2.1 hold). Let us introduce
\be
\eta(r) \defeq \widehat{E}_r(r) - \frac{Q}{r^2} \, .
\ee
Using all the assumptions we have
\begin{align}
|\eta| & = |\widehat{E}_r + 4(\LL_\FF \widehat{E}_r - \LL_\GG \widehat{B}_r)| \nonumber \\
 & \le |\widehat{E}_r| \cdot |1 + 4\LL_\FF| + 4|\LL_\GG| \cdot |\widehat{B}_r| \nonumber \\
 & \le 4(C_1 |\widehat{E}_r| + C_2 |\widehat{B}_r|) \HH \nonumber \\
 & \le 8C (|\widehat{E}_r| + |\widehat{B}_r|)(\widehat{E}_r^2 + \widehat{B}_r^2) \nonumber \\
 & \le 8C (|\widehat{E}_r| + |\widehat{B}_r|)^3 \nonumber \\
 & \le 8C \left( \frac{|Q| + |P|}{r^2} + |\eta| \right)^{\!3}
\end{align}
with $C \defeq \max\{C_1,C_2\}$. As $x \mapsto \sqrt[3]{x}$ is a monotonic function, inequality from above may be written in the following form
\be
|\eta|^{1/3} \le K \left( \frac{|Q| + |P|}{r^2} + |\eta| \right) ,
\ee
with $K \defeq 2\sqrt[3]{C}$. Assumed decrease of electromagnetic field implies $\lim_{r\to\infty}\widehat{E}_r(r) = 0$ and $\lim_{r\to\infty} \eta(r) \to 0$, thus there is $r_2 \ge r_1$, such that $|\eta| \le 1/(8K^{3/2})$ for all $r \ge r_2$. This inequality may be written, after squaring and multiplication by $|\eta|$, as $|\eta|^3 \le |\eta|/(64 K^3)$ or $|\eta| \le |\eta|^{1/3}/(4K)$. Hence,
\be
|\eta|^{1/3} \le K \left( \frac{|Q| + |P|}{r^2} + \frac{|\eta|^{1/3}}{4K} \right) ,
\ee
which leads to
\be
\frac{3}{4} \, |\eta|^{1/3} \le K \, \frac{|Q| + |P|}{r^2}
\ee
and the claim immediately follows. \qed

\medskip

As a first corollary, we may say something about the corresponding electric scalar potential $\Phi$ in a gauge where $\lim_{r\to\infty} \Phi(r) = 0$. If we introduce $\phi(r) \defeq \Phi(r) - Qr^{-1}$, then $\phi'(r) = -\eta(r)$ and for all $r \ge r_2$ there is a constant $\phi_0 > 0$, such that
\be
|\phi(r)| = \left| \int_\infty^r \phi'(s) \, \df s \right| = \left| \int_r^\infty \eta(s) \, \df s \right| \le \phi_0 \int_r^\infty \frac{\df s}{s^6} = \frac{\phi_0}{5r^5} \, .
\ee
Hence, $\Phi(r) = Qr^{-1} + O(r^{-5})$ as $r \to \infty$. An important lesson here is that the NLE corrections to the electric field and the scalar potential fall-off as, respectively, $O(r^{-6})$ and $O(r^{-5})$, given that NLE theory respects MWF limit. Such fall-off is met in the hexadecapole term of the multipole expansion in classical Maxwell's theory, thus we refer to it picturesquely as a ``hexadecapole heart'' of NLE. The analogy, however, is very limited as Maxwell's hexadecapole field is highly anisotropic, whereas the NLE field considered here is spherically symmetric. We note that this result was hinted several decades ago in \cite{Chellone71} and \cite{BS75} for test NLE fields on flat background, albeit without a rigorous proof (the authors have sketched an iterative procedure for the analytic Lagrangian). 

\smallskip

Asymptotic properties of the electric field, proven above, can be explicitly demonstrated with solutions in theories with the Born--Infeld \cite{Born34,BI34,GSP84,CK98} and Euler--Heisenberg \cite{YT00,RWX13} electromagnetic Lagrangians. On the other hand, NLE Lagrangians which do not obey MWF limit may lead to the electric field with different properties. The simplest example is the power-Maxwell Lagrangian of the form $\LL(\FF) = -\FF^p/4$; given that one chooses parameter $p = (2+\varepsilon)/(2+2\varepsilon)$ with $|\varepsilon| < 1$, spherically symmetric electric field in such a theory will have the form $\widehat{E}_r = Q r^{-2(1+\varepsilon)}$. Less trivial examples can be found among recently introduced ModMax \cite{Kosyakov20,BLST20} and RegMax \cite{HKST23} NLE theories.

\section{Complete reconstruction of the electric field}

Theorem 2.2 relies on relatively low level of differentiability of the Lagrangian density $\LL$. Let us now turn to completely opposite setting, in which we assume that $\LL$ is analytic on some neighbourhood of the origin of $\FF$-$\GG$ plane, with the Taylor series
\be
\LL(\FF,\GG) = \sum_{k,\ell = 0}^\infty c_{k\ell} \, \FF^k \GG^\ell \, .
\ee
Note that $c_{00}$ is just a constant term in Lagrangian, while the $c_{01}$ term is dynamically irrelevant, thus we may choose $c_{00} = 0$ and $c_{01} = 0$. For simplicity we look at purely electric solution ($Q \ne 0$, $P = 0$), in which $\widehat{B}_r = 0$, $\FF = -2\widehat{E}_r^2$, $\GG = 0$ and
\be
\LL_\FF(\FF,0) = \sum_{k=1}^\infty c_{k0} k \FF^{k-1} \, .
\ee
Generalized Maxwell's equations (\ref{eq:gMax}) are in this case reduced to
\be
\frac{1}{r^2} = -\frac{4}{Q} \, \LL_\FF \widehat{E}_r
\ee
and the challenge is to extract the electric field $\widehat{E}_r$ from this nonlinear equation. To this aim we turn to two mathematical tools.

\smallskip

First, Lagrange inversion theorem (see e.g.~3.6.6 in \cite{AbSt}) allows us to solve an equation of the form $z = f(w)$, with function $f$ analytic on some neighbourhood of $a$, such that $f'(a) \ne 0$, in a form of the series
\be
w = g(z) = a + \sum_{k=1}^\infty \frac{g_k}{k!} \, (z - f(a))^k \, ,
\ee
with the coefficients
\be
g_k = \lim_{w\to a} \frac{\df^{k-1}}{\df w^{k-1}} \left( \frac{w-a}{f(w) - f(a)} \right)^{\!k} \, .
\ee
Here we have $z = r^{-2}$, $a = 0$, $w = \widehat{E}_r$, $\FF = -2w^2$ and $f(w) = (-4/Q) w h(w)$, with the auxiliary function
\be
h(w) = \sum_{k=1}^\infty c_{k0} k (-2w^2)^{k-1} \, .
\ee
Hence, Lagrange inversion theorem implies
\be
\widehat{E}_r(r) = \sum_{k=1}^\infty \frac{(-Q/4)^k}{k!\,r^{2k}} \, \lim_{w\to 0} \frac{\df^{k-1}}{\df w^{k-1}} \, \frac{1}{h(w)^k} \, .
\ee
Final obstacle is the evaluation of higher derivatives under the limit, which is a straightforward, but highly tedious task. In order to gain some insight into the form of the series above, we turn to Fa\`a di Bruno's formula \cite{Johnson02,Craik05},
\be
\frac{\df^\ell}{\df w^\ell} \, F(h(w)) = \sum \frac{\ell!}{m_1! \cdots m_\ell!} \, F^{(m_1 + \cdots + m_\ell)}(h(w)) \prod_{j=1}^\ell \left( \frac{h^{(j)}(w)}{j!} \right)^{\! m_j} \, ,
\ee
where the sum is over all $\ell$-tuples of nonnegative integers $(m_1,\dots,m_\ell)$, such that $1\cdot m_1 + 2\cdot m_2 + \cdots + \ell \cdot m_\ell = \ell$. In our particular problem we choose $\ell=k-1$ and $F(x) = x^{-k}$. As $h(0) = c_{10}$ and $(k-1)! F^{(m)} = (-1)^m (k+m-1)! x^{-(k+m)}$, we have
\be
\lim_{w\to 0} (k-1)! F^{(m)}(h(w)) = (-1)^m (k+m-1)! c_{10}^{-(k+m)} \, ,
\ee
with $m \defeq m_1 + \cdots + m_{k-1}$. Furthermore, note that $h^{(2p-2)}(0) = (2p-2)! p(-2)^{p-1} c_{p 0}$ and $h^{(2p+1)}(0) = 0$. Hence, the product at the end of Fa\`a di Bruno's formula is zero whenever there is nonzero $m_j$ with odd index $j$. In other words, the only nontrivial contributions to Fa\`a di Bruno's sum are those in which all $m_j$ with odd index $j$ are zero. In such cases $k-1$ is even, i.e.~$k$ is odd. If we introduce new index via $k=2n-1$, the electric field series may be written in a compact form 
\be\label{eq:seriesE}
\widehat{E}_r(r) = \sum_{n=1}^\infty \frac{e_n Q^{2n-1}}{r^{2(2n-1)}} \, ,
\ee
with all the ``clutter'' packed in coefficients $e_n$. It is easy to see that $e_1 = 1$, but form of the higher coefficients quickly grows in complexity (see table 1). One can immediately notice that the first correction to the Maxwell's $n=1$ term is of the order $O(r^{-6})$, in agreement with the theorem 2.2.

\bc
\begin{table}[ht]
\centering
\begin{tabular}{rl}
$n$ & $e_n$ \\
\hline
$1$ & $1$ \\
$2$ & $-16 \, c_{20}$ \\
$3$ & $48 \, (16 \, c_{20}^2 + c_{30})$ \\
$4$ & $-128 \, (384 \, c_{20}^3 + 48 \, c_{20} c_{30} + c_{40})$ \\
$5$ & $320 \, (11\,264 \, c_{20}^4 + 2112 \, c_{20}^2 c_{30} + 64 \, c_{20} c_{40} + 36 \, c_{30}^2  + c_{50})$
\end{tabular}

\caption{First five coefficients $e_n$ in the expansion of the electric field $\widehat{E}_r$.}
\end{table}
\ec

\section{Final remarks}

We have shown how exactly behaviour of the NLE Lagrangian near the origin of the $\FF$-$\GG$ plane controls the asymptotic form of the electric field. Heuristically, it is not a surprise that Lagrangian terms with higher powers of electromagnetic invariants reflect in higher power terms in the expansion of the electromagnetic field, but it takes a bit of rigorous treatment to translate this intuition into a precise, quantitative statement. An obvious next step would be to generalize theorem 2.2 to the stationary, not necessarily spherically symmetric spacetimes, and one can expect that the result will be essentially similar to the static case. This problem is closely related to the question of gyromagnetic ratio \cite{Wald74,PK02} for the charged, rotating black holes with nonlinear electromagnetic fields \cite{CK98}.

\smallskip 

Reconstruction of the electric field from the section 3 is, in general, only formal, as the electric field 1-form $\pf{E}$ and consequently the electromagnetic 2-form $\pf{F}$ depend on the product $\alpha(r)\beta(r)$, which is restricted by the gravitational field equation, which in turn contains the electromagnetic field. However, as was already emphasized above, Einstein's gravitational field equation implies that $\alpha(r)\beta(r)$ is constant and in that case the series (\ref{eq:seriesE}) directly leads to the electric field 1-form $\pf{E}$.

\ack
The research was supported by the Croatian Science Foundation Project No.~IP-2020-02-9614.


\section*{References}

\bibliographystyle{iopnum}
\bibliography{nleasym}

\end{document}